
\input phyzzx.tex
\input psfig.tex
\hoffset=1cm
\voffset=1pc
\overfullrule0pt
\hsize=35.5pc \vsize=51pc
\normalparskip=0pt
\parindent=36pt
\itemsize=36pt
\def\linebreak{\unskip\break}
\def\ifmath#1{\relax\ifmmode #1\else $#1$\fi}
\def\br{{\rm BR}}
\def\rta{\rightarrow}
\def\wp{W^+}
\def\wm{W^-}
\def\mt{m_t}
\def\tanb{\tan\beta}

\def\hl{h^0}
\def\mhl{m_{\hl}}
\def\ha{A^0}
\def\mha{m_{\ha}}
\def\hh{H^0}
\def\mhh{m_{\hh}}
\def\hpm{H^{\pm}}

\def\sb{s_{\beta}}
\def\cb{c_{\beta}}

\def\mw{m_W}
\def\mz{m_Z}
\def\hpm{H^{\pm}}
\def\mhpm{m_{\hpm}}

\def\hp{H^+}
\def\hm{H^-}

\def\mlsq{m\ls{L}^2}
\def\mtsq{m\ls{T}^2}
\def\mdsq{m\ls{D}^2}
\def\mssq{m\ls{S}^2}
\def\calm{{\cal M}}
\def\calv{{\cal V}}
\def\crrr{\cr\noalign{\vskip8pt}}

\def\GENITEM#1;#2{\par\vskip6pt \hangafter=0 \hangindent=#1
   \Textindent{$ #2$ }\ignorespaces}

\def\unlock{\catcode`@=11} 
\def\lock{\catcode`@=12} 
\unlock
\def\ls#1{_{\lower1.5pt\hbox{$\scriptstyle #1$}}}
\def\chapter#1{\par \penalty-300 \vskip\chapterskip
   \spacecheck\chapterminspace
   \chapterreset \leftline{\bf \chapterlabel.~~#1}
   \nobreak\vskip\headskip \penalty 30000
   {\pr@tect\wlog{\string\chapter\space \chapterlabel}} }
\def\section#1{\par \ifnum\lastpenalty=30000\else
   \penalty-200\vskip\sectionskip \spacecheck\sectionminspace\fi
   \gl@bal\advance\sectionnumber by 1
   {\pr@tect
   \xdef\sectionlabel{\chapterlabel.%
       \the\sectionstyle{\the\sectionnumber}}%
   \wlog{\string\section\space \sectionlabel}}%
   \noindent {\it\sectionlabel.~~#1}\par
   \nobreak\vskip\headskip \penalty 30000 }\lock
\def\CERN{\centerline{\it CERN, Theory Division}
  \centerline{CH-1211 Geneva 23, Switzerland}}

\Pubnum={CERN-TH/95-109 \cr
SCIPP 95/15}
\date={May, 1995}
\titlepage
\singlespace
\vbox to 1cm{}
\centerline{{\fourteenbf  CHALLENGES FOR NON-MINIMAL HIGGS SEARCHES}}
\vskip4pt
\centerline{{\fourteenbf  AT FUTURE COLLIDERS%
\foot{Work supported in part by the
U.S.~Department of Energy.}}}
\vskip1cm
\centerline{{\caps Howard E. Haber}%
\foot{Permanent address: Santa Cruz Institute for Particle Physics,
University of California, Santa Cruz, CA 95064 USA.}}
\vskip .1in
\CERN
\vskip1cm
\centerline{\bf Abstract}
\vskip6pt
In models with a non-minimal Higgs sector, the lightest scalar state
may be a neutral CP-even Higgs boson, whose properties are
nearly identical to those of the minimal Higgs boson of the
Standard Model.  In such a scenario, the other Higgs scalars
are significantly heavier than $m_Z$; their effects
rapidly decouple from the low-energy theory.
The decoupling limit of the most general CP-conserving
two-Higgs doublet model is formulated.   Detection of evidence for a
non-minimal Higgs sector in the decoupling limit presents a
formidable challenge for Higgs searches at future colliders.
\vfill
\centerline{Invited Talks presented at the}
\centerline{Fourth International Conference on Physics
Beyond the Standard Model,}
\centerline{Lake Tahoe, CA, USA,  13--18 December, 1994, }
\centerline{and at the}
\centerline{Workshop on {\it Perspectives for Electroweak Interactions in}
$e^+e^-$ {\it Collisions},}
\centerline{Ringberg Castle, Tegernsee, Germany, 5--8 February, 1995.}
\vfill
\endpage
\medskip
\chapter{Introduction}
\medskip

\REF\hhg{J.F. Gunion, H.E. Haber, G.L. Kane and S. Dawson,
{\it The Higgs Hunter's Guide} (Addison-Wesley, Redwood City, CA, 1990).}
\REF\cpr{P.H. Chankowski, S. Pokorski and J. Rosiek, {\sl Phys. Lett.}
{\bf B281} (1992) 100.}
\REF\gkane{G.L. Kane, in {\it Perspectives on Higgs Physics}, edited by
G.L. Kane (World Scientific, Singapore, 1993) pp.~223--228.}
Let us suppose that a candidate for the Higgs boson is discovered in
a future collider experiment.  What are the expectations for its
properties?\refmark\hhg\
Will it resemble the Higgs boson of the minimal Standard
Model or will it possess some distinguishing trait?  If the
properties of this scalar state are difficult to distinguish from the
Standard Model Higgs boson, what are the requirements of future
collider experiments for detecting the existence (or non-existence)
of a non-minimal Higgs sector?
See \eg, refs.~\cpr\ and \gkane\ for earlier attempts to address these
questions..

I shall address these questions in the context of the
(CP-conserving)
two-Higgs doublet model.  In this model, the physical scalar states
consist of a charged Higgs pair ($\hpm$), two CP-even scalars ($\hl$
and $\hh$, with $\mhl\le\mhh$) and one CP-odd scalar ($\ha$).
The ultimate conclusions of this paper will
survive in models with more complicated scalar sectors.
The working hypothesis of this
paper is that $\hl$, assumed to be the lightest scalar state,
will be the first Higgs boson to be discovered.
Moreover,
the mass gap between $\hl$ and the heavier scalars is assumed to be
sufficiently large so that the initial experiments which can detect $\hl$
will not have sufficient energy and luminosity to discover
any of the heavier scalar states.

\REF\pdg{L. Montanet \etal\ [Particle Data Group] {\sl Phys. Rev.}
{\bf D50} (1994) 1173.}
\REF\sopczak{A. Sopczak, {\sl Int. J. Mod. Phys.} {\bf A9}
(1994) 1747.}
\REF\mrenna{A. Stange, W. Marciano and S. Willenbrock, {\sl Phys. Rev.}
{\bf D49} (1994) 1354; {\bf D50} (1994) 4491; S. Mrenna and G.L. Kane,
CALT-68-1938 (1994) [hep-ph 9406337].}
\REF\zeuthen{J.F. Gunion,
{\sl Nucl. Phys. B} (Proc. Suppl.) {\bf 37B} (1994) 253.}
\REF\heavyhiggs{D. Froidevaux, Z. Kunszt and J. Stirling \etal, in
{\it Proceedings of the Large Hadron Collider Workshop},
Aachen 1990, CERN Report 90-10 (1990).}
\REF\imh{S. Dawson, in {\it Perspectives on Higgs Physics}, edited
by G.L. Kane (World Scientific, Singapore, 1993) pp.~129--155; Z. Kunszt,
{\it ibid.}~pp.~156--178.}
\REF\bargeretal{V. Barger, K. Cheung, A. Djouadi, B.A. Kniehl, and
P. Zerwas, {\sl Phys. Rev.} {\bf D49} (1994) 79.}
\REF\ghphoton{J.F. Gunion and H.E. Haber,
{\sl Phys. Rev.} {\bf D48} (1993) 5109.}
\REF\caldwell{D.L. Borden, D.A. Bauer and D.O. Caldwell,
{\sl Phys. Rev.} {\bf D48} (1993) 4018.}
How will $\hl$ be discovered and where?  Present LEP bounds\refmark\pdg\
imply
that $\mhl\gsim 60$~GeV.  This bound will be improved by
LEP-II,\refmark\sopczak\
which will be sensitive to Higgs masses up to roughly $\sqrt{s}-\mz-
10$~GeV.  The LEP search is based on $e^+e^-\rta Z\rta Z\hl$ where
one of the two $Z$'s is on-shell and the other is off-shell.   At hadron
colliders, an upgraded Tevatron with an integrated luminosity of
10~fb$^{-1}$ can begin to explore the intermediate-mass Higgs
regime\refmark{\mrenna,\zeuthen}
($80\lsim\mhl\lsim 130$~GeV).  The Higgs search at the LHC will
significantly extend the Higgs search to higher
masses\refmark{\zeuthen,\heavyhiggs}
(although the intermediate mass regime still presents some significant
difficulties for the LHC
detector collaborations).  The dominant mechanism for Higgs
production at hadron colliders is via $gg$-fusion through a top-quark
loop.  If $\mhl>2\mz$, the ``gold-plated'' detection mode is $\hl\rta
ZZ$; each $Z$ subsequently decays leptonically, $Z\rta\ell^+\ell^-$
(for $\mhl\gsim 130$~GeV, $\hl\rta ZZ^\ast$, where $Z^\ast$ is
off-shell, provides a viable signature).  Other decay modes are required
in the case of the intermediate mass Higgs
(for recent reviews, see refs.~\zeuthen\ and \imh).
At a future $e^+e^-$ linear collider (NLC), the Higgs mass
reach of LEP-II will be extended.\refmark\bargeretal\  In addition,
with increasing $\sqrt{s}$, Higgs boson production via
$W^+W^ -$ fusion begins to be the dominant production process.
Finally,  Higgs production via $\gamma\gamma$-fusion through a
$\wp\wm$ and a $t\bar t$ loop may be detectable at the $\gamma\gamma$
collider option of the NLC,
depending on the particular Higgs final
state decay.\refmark{\ghphoton, \caldwell}\
Note that almost all of the Higgs search techniques outlined
above involve the $\hl ZZ$ (and in some cases the
$\hl\wp\wm$) vertex.   In a few cases, it is the $\hl t\bar t$ vertex
(and possibly the $\hl b\bar b$ vertex) that plays the key role.

In section 2, I briefly review the Higgs sector of the minimal supersymmetric
extension of the Standard Model (MSSM).  In this context, I discuss
under which circumstances
one might expect $\hl$ to be the lightest scalar whose
properties are nearly identical to those of the Standard Model Higgs
boson.  In section 3, I place the results of section 2 in a more
general context.  I define the ``decoupling limit'' of the general
two-Higgs doublet model; in this limit, $\hl$ is indistinguishable from
the Standard Model Higgs boson.  In section 4, I discuss the
phenomenological challenges of the decoupling limit for the Higgs search
at future
colliders.  After briefly mentioning the prospects for non-minimal
Higgs detection at the LHC, I consider in more detail the prospects
for the discovery of the non-minimal Higgs sector at the NLC.
Conclusions are presented in section 5.

\medskip
\chapter{The Higgs Sector of the MSSM---A Brief Review}
\medskip

\REF\hhgsusy{J.F. Gunion and H.E. Haber, {\sl Nucl. Phys.} {\bf B272}
(1986) 1; {\bf B278} (1986) 449 [E: {\bf B402} (1993) 567].}
The Higgs sector of
the minimal supersymmetric extension of the Standard Model
(MSSM) consists of two complex doublet scalar fields $H_1$ and $H_2$
of hypercharge $-1$ and $+1$, respectively.\refmark\hhgsusy\
Because of the underlying supersymmetry, the tree-level Higgs masses
and couplings are determined in terms of two free parameters:
$\mha$ and $\tanb=v_2/v_1$ [where $v_2$ ($v_1$) is the vacuum expectation
value of the Higgs field that couples to up-type (down-type) fermions].
Then, the other (tree-level) Higgs masses are given by
$$\eqalign{
  m^2_{H^\pm} &= m^2_W +\mha^2\,,\cr
  m^2_{H^0,\,h^0} &= \half \left(\mha^2+m^2_Z \pm
       \sqrt{(\mha^2+m^2_Z)^2 - 4m^2_Z \mha^2 \cos^2 2\beta}\right)\,.\cr}
\eqn\treemasses
$$
The mass eigenstates $H^0$ and $h^0$ are linear combinations of
the original Higgs fields of definite hypercharge
$$\eqalign{H^0 &= (\sqrt 2\,{\rm Re}\,H^0_1-v_1) \cos\alpha +
(\sqrt2\,{\rm Re}\,H^0_2-v_2)
              \sin\alpha   \cr
  h^0 &= -(\sqrt 2\,{\rm Re}\,H^0_1-v_1)\sin\alpha + (\sqrt2\,{\rm Re}
\,H^0_2-v_2)
             \cos\alpha \cr
}\eqn\cpevenhiggs$$
which defines the CP-even Higgs mixing angle $\alpha$.
Explicit formulae for $\alpha$ can also
be derived.  Here, I shall note one particularly useful relation
$$\cos^2(\beta-\alpha)={\mhl^2(\mz^2-\mhl^2)\over
\mha^2(\mhh^2-\mhl^2)}\,.\eqn\cbmasq$$

\REF\impact{For a review of the influence of radiative corrections on the
MSSM Higgs sector, see H.E. Haber, in {\it Perspectives on Higgs
Physics}, edited by G.L. Kane (World Scientific, Singapore, 1993) pp.~79--128.}

Consider the limit where $\mha\gg\mz$.  Then, from the above
formulae,
$$\eqalign{
\mhl^2\simeq\ &\mz^2\cos^2 2\beta\,,\cr
\mhh^2\simeq\ &\mha^2+\mz^2\sin^2 2\beta\,,\cr
\mhpm^2=\ & \mha^2+\mw^2\,,\cr
\cos^2(\beta-\alpha)\simeq\ &{\mz^4\sin^2 4\beta\over 4\mha^4}\,.\cr}
\eqn\largema$$Two consequences are immediately apparent.
First, $\mha\simeq\mhh
\simeq\mhpm$, up to corrections of ${\cal O}(\mz^2/\mha)$.  Second,
$\cos(\beta-\alpha)=0$ up to corrections of ${\cal O}(\mz^2/\mha^2)$.
Although the radiative corrections to the Higgs masses
can have a profound effect on the phenomenology,\refmark\impact\
the overall size of
such corrections is never larger than ${\cal O}(\mz)$, and hence
the consequences of eq.~\largema\ noted above remain valid.

\REF\wudka{J.F. Gunion, H.E. Haber, and J. Wudka, {\sl Phys. Rev.}
{\bf D43} (1991) 904.}
The phenomenological implications of these results may be discerned
by reviewing the coupling strengths of the Higgs bosons to
Standard Model particles (gauge bosons, quarks and leptons) in the
two-Higgs doublet model.  The coupling of $\hl$ and $\hh$ to vector
boson pairs or vector-scalar boson final states is proportional
%
%
to either $\sin(\beta-\alpha)$ or $\cos(\beta-\alpha)$ as indicated
below.\refmark{\hhg,\hhgsusy}

\vskip 0.2in
\settabs 6 \columns
\+&\us{$\cos(\beta-\alpha)$}&&&\us{$\sin(\beta-\alpha)$}\cr
\vskip 0.1in\+&$\hh\wp\wm$&&&$\hl\wp\wm$\cr
\+&$\hh ZZ$&&&$\hl ZZ$\cr
\+&$Z\ha\hl$&&&$Z\ha\hh$\cr
\+&$W^\pm H^\mp\hl$&&&$W^\pm H^\mp\hh$\cr
\+&$ZW^\pm H^\mp\hl$&&&$ZW^\pm H^\mp\hh$\cr
\+&$\gamma W^\pm H^\mp\hl$&&&$\gamma W^\pm H^\mp\hh$\cr
\vskip 0.2in\noindent
Note in particular that {\it all} vertices
in the theory that contain at least
one vector boson and {\it exactly one} heavy Higgs boson state
($\hh$, $\ha$ or $\hpm$) are proportional to $\cos(\beta-\alpha)$.
This can be understood as a consequence of unitarity sum rules which
must be satisfied by the tree-level amplitudes of the theory.\refmark\wudka\

\def\phm{\phantom{-}}
In models with a non-minimal Higgs sector,
the Higgs couplings to quarks and leptons are model-dependent.
In the MSSM, one Higgs doublet ($H_1$) couples
exclusively to down-type fermions and the second Higgs doublet ($H_2$)
couples exclusively to up-type fermions.  In this case, the couplings
of the neutral CP-even Higgs bosons to $f\bar f$ relative to the
corresponding Standard Model value (using 3rd family notation) are given by
$$\eqalign{\hl b\bar b:\qquad -{\sin\alpha\over\cos\beta}=\ &\sin(\beta-\alpha)
-\tan\beta\cos(\beta-\alpha)\,,\crr
\hl t\bar t:\qquad \phm{\cos\alpha\over\sin\beta}=\ &\sin(\beta-\alpha)
+\cot\beta\cos(\beta-\alpha)\,,\crr
\hh b\bar b:\qquad \phm{\cos\alpha\over\cos\beta}=\ &\cos(\beta-\alpha)
+\tan\beta\sin(\beta-\alpha)\,,\crr
\hh t\bar t:\qquad \phm{\sin\alpha\over\sin\beta}=\ &\cos(\beta-\alpha)
-\cot\beta\sin(\beta-\alpha)\,.\cr}\eqn\hffcoup$$
In contrast to the Higgs couplings to vector bosons,
none of the couplings in eq.~\hffcoup\ vanish when $\cos(\beta-
\alpha)=0$.
The significance of $\cos(\beta-\alpha)= 0$ is now evident: in this limit,
couplings of $\hl$ to
gauge boson pairs and fermion pairs are identical to the couplings of
the Higgs boson in the minimal Standard Model.\foot{Likewise, the
$\hl\hl\hl$ and $\hl\hl\hl\hl$ couplings also reduce to their Standard Model
values when $\cos(\beta-\alpha)=0$, while in the same limit the
other Higgs self-couplings (which involve $\hh$, $\ha$, and/or $H^\pm$)
do {\it not} vanish.}
More precisely, in
the limit of $\mha\gg\mz$, the effects of the heavy Higgs states
($\hpm$, $\hh$ and $\ha$) decouple, and the low-energy effective
scalar sector is indistinguishable from that of the minimal
Standard Model.

\REF\thomas{H.E. Haber and S. Thomas, SCIPP preprint in preparation.}
\REF\nir{H.E. Haber and Y. Nir, {\sl Nucl. Phys.} {\bf B335}
(1990) 363.}

\medskip
\chapter{Decoupling Properties of the Two-Higgs Doublet Model\refmark\thomas}
\medskip

The decoupling properties of the MSSM Higgs sector are not special to
supersymmetry.  Rather, they are a generic feature of non-minimal
Higgs sectors.\refmark\nir\  In this section, I demonstrate this
assertion in the case of the most general CP-conserving two-Higgs doublet
model.
Let $\Phi_1$ and
$\Phi_2$ denote two complex $Y=1$, SU(2)$\ls{L}$ doublet scalar
fields.\foot{In terms of the $Y=\pm1$ fields of the previous section,
$H_1^i=\epsilon_{ij}{\Phi_1^j}^\star$ and $H_2=\Phi_2$.}
The most general gauge invariant scalar potential is given by
$$\eqalign{
\calv&=m_{11}^2\Phi_1^\dagger\Phi_1+m_{22}^2\Phi_2^\dagger\Phi_2
-[m_{12}^2\Phi_1^\dagger\Phi_2+{\rm h.c.}]\crrr
&\quad +\half\lambda_1(\Phi_1^\dagger\Phi_1)^2
+\half\lambda_2(\Phi_2^\dagger\Phi_2)^2
+\lambda_3(\Phi_1^\dagger\Phi_1)(\Phi_2^\dagger\Phi_2)
+\lambda_4(\Phi_1^\dagger\Phi_2)(\Phi_2^\dagger\Phi_1)\crrr
&\quad +\left\{\half\lambda_5(\Phi_1^\dagger\Phi_2)^2
+\big[\lambda_6(\Phi_1^\dagger\Phi_1)
+\lambda_7(\Phi_2^\dagger\Phi_2)\big]
\Phi_1^\dagger\Phi_2+{\rm h.c.}\right\}\,.\cr}\eqn\pot$$\vskip5pt\noindent
In principle, $m_{12}^2$, $\lambda_5$,
$\lambda_6$ and $\lambda_7$ can be complex.  Here, I shall
ignore the possibility of CP-violating effects in the Higgs sector
by choosing all coefficients in eq.~\pot\ to be real.
The scalar fields will
develop non-zero vacuum expectation values if the mass matrix
$m_{ij}^2$ has at least one negative eigenvalue. Imposing CP invariance
and U(1)$\ls{\rm EM}$ gauge symmetry, the minimum of the potential is
$\VEV{\Phi_i^0}\equiv v_i/\sqrt{2}$ ($i=1,2$),
where the $v_i$ can be chosen to be real and positive.
It is convenient to introduce the following notation:
$v^2\equiv v_1^2+v_2^2=(246~{\rm GeV})^2$
and $t_\beta\equiv\tanb\equiv v_2/v_1$.
Eliminating $m_{11}^2$ and $m_{22}^2$ by
minimizing the scalar potential, the
squared masses for the CP-odd and charged Higgs states are obtained:
$$\eqalign{%
\mha^2 &={m_{12}^2\over \sb\cb}-\half
v^2\big(2\lambda_5+\lambda_6 t_\beta^{-1}+\lambda_7t_\beta\big)\,,\cr
m_{H^{\pm}}^2 &=m_{A^0}^2+\half v^2(\lambda_5-\lambda_4)\,,
\cr}\eqn\generalmasses$$
where $s_\beta\equiv\sin\beta$ and $c_\beta\equiv\cos\beta$.
The two CP-even Higgs states mix according to the following squared mass
matrix:
$$
\calm^2 =m_{A^0}^2 \left(\matrix{\sb^2&-\sb\cb\cr
-\sb\cb&\cb^2}\right)+v^2\left(\matrix{\calm^2_{11}&\calm^2_{12}\cr
\calm^2_{12}&\calm^2_{22}\cr}\right)\,,\eqn\massmhh$$
where
$$\left(\matrix{\calm^2_{11}&\calm^2_{12}\cr
\calm^2_{12}&\calm^2_{22}\cr}\right)\equiv
\left( \matrix{\lambda_1\cb^2+2\lambda_6\sb\cb+\lambda_5\sb^2
&(\lambda_3+\lambda_4)\sb\cb+\lambda_6
\cb^2+\lambda_7\sb^2\crr
(\lambda_3+\lambda_4)\sb\cb+\lambda_6
\cb^2+\lambda_7\sb^2&
\lambda_2\sb^2+2\lambda_7\sb\cb+\lambda_5\cb^2}\right)
\eqn\calmdef$$
It is convenient to define four squared mass combinations:
$$\eqalign{\mlsq\equiv&\ \calm^2_{11}\cos^2\beta+\calm^2_{22}\sin^2\beta
+\calm^2_{12}\sin2\beta\,,\crr
\mdsq\equiv&\ \left(\calm^2_{11}\calm^2_{22}-\calm^4_{12}\right)^{1/2}\,,\crr
\mtsq\equiv&\ \calm^2_{11}+\calm^2_{22}\,,\crr
\mssq\equiv&\ \mha^2+\mtsq\,.\cr}
\eqn\massdefs$$
In terms of these quantities,
$$ m^2_{\hh,\hl}=\half\left[\mssq\pm\sqrt{m\ls{S}^4-4\mha^2\mlsq
-4m\ls{D}^4}\,\right]\,,\eqn\cpevenhiggsmasses$$
and
$$\cos^2(\beta-\alpha)= {\mlsq-\mhl^2\over\mhh^2-\mhl^2}\,.
\eqn\cosbmasq$$

Suppose that all the Higgs self-coupling constants $\lambda_i$ are
held fixed such that $\lambda_i\lsim1$, while taking
$\mha^2\gg\lambda_iv^2$.
This will be called the {\it decoupling limit} of the model.
Then the $\calm^2_{ij}\sim{\cal O}(v^2)$, and
it follows that:
$$\mhl\simeq m\ls{L}\,,\qquad\qquad \mhh\simeq\mha\simeq\mhpm\,,
\eqn\approxmasses$$
and
$$\cos^2(\beta-\alpha)\simeq\, {\mlsq(\mtsq-\mlsq)-m\ls{D}^4\over\mha^4}\,.
\eqn\approxcosbmasq$$
Comparing these results with those of eq.~\largema, one sees that the
MSSM results are simply a special case of the more general
two-Higgs doublet model results just obtained.
In particular, eq.~\approxcosbmasq\ implies that $\cos(\beta-\alpha)=
{\cal O}(\mz^2/\mha^2)$ in the decoupling limit, which means that
the $\hl$ couplings to
Standard Model particles match precisely those of the Standard Model
Higgs boson.

It is interesting to attempt to circumvent the decoupling limit
while maintaining the hierarchy of Higgs masses,
$\mhl\ll\mhh,\mha,\mhpm$.  The latter implies [using
eq.~\cpevenhiggsmasses] that
$$0<\mha^2\mlsq+m\ls{D}^4\ll m\ls{S}^4\,.\eqn\massinequality$$
Eq.~\massinequality\ is satisfied in one of two cases:\item{(i)}$\mz^2$,
$\mlsq$, $m\ls{D}^4/\mha^2\ll \mha^2$, $\mssq$.
That is, each term on the left-hand side of eq.~\massinequality\ is
separately smaller in magnitude than $m\ls{S}^4$, or
\item{(ii)} $\mha^2\mlsq$ and $m\ls{D}^4$ are both of ${\cal O}(m\ls{S}^4)$,
but due to cancelation of the leading behavior of each term, the sum
satisfies eq.~\massinequality.

\noindent In case (i), one finds that
$$\mhl^2\simeq {\mha^2\mlsq\over\mssq}+{m\ls{D}^4\over\mssq}
+{\mha^4m\ls{L}^4\over m\ls{S}^6}+{\cal O}\left({m\ls{L}^4\over
m\ls{S}^4}\right)\,,\eqn\mhlcasei$$
and $\mhh^2\sim{\cal O}(\mssq)$.  In the same approximation,
$$\cos^2(\beta-\alpha)\simeq{\mlsq\over\mssq}\left(1-
{\mha^2\over\mssq}\right)+{1\over m\ls{S}^4} \left[m\ls{L}^4\left(
{2\mha^2\over\mssq}-{3\mha^4\over m\ls{S}^4}\right)
-m\ls{D}^4\right]\,.
\eqn\cbmacasei$$
The behavior of $\cos(\beta-\alpha)$ depends crucially on how close
$\mha^2/\mssq$ is to 1.  If   \linebreak
$\mtsq\ll\mssq$ [see eq.~\massdefs],
we recover the results of the decoupling limit [see eqs.~\approxmasses\
and \approxcosbmasq].  On the other
hand, if $\mtsq\sim{\cal O}(\mssq)$, then eq.~\cbmacasei\ implies
that $\cos(\beta-\alpha)\sim{\cal O}(\mz/\mha)$.  This is a particular
region of the parameter space where some of the $\lambda_i$ are
substantially larger than 1, and yet the $\hl$ couplings do not
significantly deviate from those of the Standard Model.
Nevertheless, the onset of decoupling is slower than the
$\cos(\beta-\alpha)\sim{\cal O}(\mz^2/\mha^2)$ behavior found in the
decoupling regime.   In order to find a parameter regime which
exhibits complete non-decoupling, one must consider case (ii) above.
In this case, despite the fact that $\mhl\ll\mhh,\mha,\mhpm$, one
nevertheless finds that $\cos(\beta-\alpha)\sim{\cal O}(1)$, which implies
that the couplings of $\hl$ deviate significantly from those of the
Standard Model Higgs boson.  Although it might appear that case (ii)
requires an unnatural cancelation, it is easy to construct a simple
model of non-decoupling.  Consider a model where:
$m_{12}^2=\lambda_6=\lambda_7=0$, $\lambda_3+\lambda_4+\lambda_5=0$,
$\lambda_2\lsim{\cal O}(1)$, and $\lambda_1, \lambda_3, -\lambda_5\gg1$.
This model yields: $\mhl^2=\lambda_2 v^2 s_\beta^2$,
$\mhh^2=\lambda_1 v^2
c_\beta^2$, $\mha^2=-\lambda_5 v^2$, $\mhpm^2=\half\lambda_3 v^2$,
and $\cos^2(\beta-\alpha)=c_\beta^2$.  Note that in this model,
the heavy Higgs states are {\it not}
approximately degenerate (as required in the decoupling limit).

\chapter{Phenomenological Challenges of the Decoupling Limit}
\medskip

In the decoupling limit, the couplings of $\hl$ to
Standard Model gauge bosons and fermions approach those of the
Standard Model Higgs boson.  Suppose that a future experiment has
already discovered and studied the properties of $\hl$.
What are the requirements of experiments at future colliders
for proving the existence or non-existence of a non-minimal Higgs sector?
To be specific, let us assume in this section that $\hl$ has been
discovered with couplings approximating those of the Standard Model
Higgs boson and $\mha>250$~GeV.

\REF\xsecs{For a recent review, see
J.F. Gunion, in {\it Perspectives on Higgs Physics},
edited by G.L. Kane (World Scientific, Singapore, 1993) pp.~179--222.}
\REF\higgstt{D. Dicus, A. Stange and S. Willenbrock, {\sl Phys. Lett.}
{\bf B333} (1994) 126.}
\REF\vega{J. Dai, J.F. Gunion and R. Vega, {\sl Phys. Rev. Lett.}
{\bf 71} (1993) 2699; {\sl Phys. Lett.} {\bf B345} (1995) 29.}
At the LHC, the rate for $gg\rta \ha$, $\hh$ and $gb\rta\hm t$ provides
a substantial number of produced Higgs bosons per year (assuming
that the heavy Higgs masses are not too large).\refmark\xsecs\
Unfortunately, there may not be a viable final state signature.  For
example, since $\cos^2(\beta-\alpha)\ll 1$, the branching ratio
of $\hh\rta ZZ$ is significantly suppressed, so that the gold-plated
Standard Model Higgs signature is simply absent.  At present, there
is no known proven technique for detecting $\ha$, $\hh$ and $\hpm$ signals
at the LHC in the decoupling regime of parameter space.  An
attempt to isolate a Higgs signal in $t\bar t$ final states has been
discussed in ref.~\higgstt.  Another method consists of a search for Higgs
signals in $t\bar tt\bar t$, $t\bar t b\bar b$, and $b\bar bb\bar b$
final states.\refmark\vega\
These can arise from $gg\rta Q\bar Q^\prime(\hh$, $\ha$ or
$\hpm$), where $Q$ is a heavy quark ($b$ or $t$), and the Higgs boson
subsequently decays into a heavy quark pair.  As noted
below eq.~\hffcoup, even in the decoupling
limit, the couplings of $\hh$, $\ha$ and $\hpm$ to heavy quark pairs
are not suppressed.  Whether such signals can be extracted from the
substantial QCD backgrounds (very efficient $b$-tagging is one of the
many requirements for such a search) remains to be seen.\foot{For an optimistic
assessment, see ref.~\zeuthen.  In the large $\tanb$
regime, the enhanced couplings of the heavy Higgs bosons to $b\bar b$
could lead to an observable signal in $gg\rta b\bar b(\hh$ or
$\ha$) followed by Higgs decay to either $b\bar b$ or $\tau^+\tau^-$.}

\REF\janot{P. Janot, in
{\it Physics and Experiments with Linear $\epem$ Colliders},
Workshop Proceedings, Waikoloa, Hawaii, 26--30 April, 1993,
edited by F.A. Harris, S.L. Olsen, S. Pakvasa and X. Tata
(World Scientific, Singapore, 1993) pp.~192--217.}
\REF\nlcsearch{A. Djouadi, J. Kalinowski and P.M. Zerwas, {\sl Z. Phys.}
{\bf C57} (1993) 569.}
\REF\djouadi{A. Djouadi, J. Kalinowski and P.M. Zerwas,
{\sl Mod. Phys. Lett.} {\bf A7} (1992) 1765; {\sl Z. Phys.} {\bf C54}
(1992) 255.}
Let us now turn to $e^+e^-$ colliders.  Consider the process
$e^+e^-\rta\hl\ha$ (via $s$-channel $Z$-exchange).  Since the $Z\hl\ha$
coupling is proportional to $\cos(\beta-\alpha)$, the
production rate is suppressed in the decoupling regime.  For example,
in the MSSM, if $\mha>\mhl$, then LEP-II will not
possess sufficient energy and/or luminosity to directly produce the
$\ha$.\refmark\janot\  Of course, with sufficient energy, one can directly
pair-produce the heavy Higgs states via $e^+e^-\rta H^+H^-$ and
$e^+e^-\rta\hh\ha$ without a rate suppression.
At the NLC, with $\sqrt{s}=500$~GeV and 10~ fb$^{-1}$ of
data, it has been shown\refmark{\janot,\nlcsearch}
that no direct signals for $\ha$, $\hh$, and
$\hpm$ can be seen if $\mha\gsim 230$~GeV.
These results support the
following general conclusion: {\it evidence for the non-minimal
Higgs sector at the NLC requires a machine with energy
$\sqrt{s}>2\mha$, sufficient to pair-produce heavy Higgs states}.
Can this conclusion be avoided?  There are two possible methods:
(i) produce one heavy Higgs state in association with
lighter states, or
(ii) make precision measurements of $\hl$ couplings to
Standard Model particles in order to detect a deviation from the
Standard Model expectations.

First, consider the processes that yield a singly produced
heavy Higgs state.  In section 2, I noted that whenever a single
heavy Higgs state couples directly to a gauge boson plus other light
particles, the coupling is suppressed by $\cos(\beta-\alpha)$.
To avoid this suppression, one must couple the heavy Higgs state to
either fermions or scalars.  For example, the production rates for
$e^+e^-\rta Q\bar
Q^\prime(\hh,\ha$, or $\hpm)$, where $Q=b$ or $t$,
have been computed by Djouadi \etal\refmark\djouadi\
Unfortunately, the three-body
phase space greatly suppresses the rate once the heavy Higgs state
is more massive than the $Z$.  In particular, for
$\mha>\sqrt{s}/2$, these processes do not provide viable signatures
for the heavy Higgs states.  A similar conclusion is
obtained when the heavy Higgs state couples to light scalars.
Thomas and I have computed the rate for $e^+e^-\rta\hl\hh Z$, assuming
that the dominant contribution is due to the $s$-channel $Z$-exchange,
where the virtual $Z^\ast$ decays via $Z^\ast\rta Z{\hl}^\ast\rta Z\hl\hh$.
For $\mhh\gg\mhl,\mz$, we obtained\refmark\thomas\
$$\eqalign{
{\sigma(e^+e^-\rta\hl\hh Z)\over\sigma(e^+e^-\rta\hl Z)}\simeq
{g^2_{\hh\hl\hl}\over 32\pi^2s^3\mhh^2}\biggl\{(s-\mhh^2)&\left[(s-\mhh^2)^2
+12s\mhh^2\right]\cr
-6\mhh^2& s(s+\mhh^2)\ln\left({s\over\mhh^2}\right)\biggr\}\,,\cr}
\eqn\threeh$$
where the $\hh\hl\hl$ coupling in the decoupling limit
[{\it i.e.}, when $\cos(\beta-\alpha)=0$] in the MSSM is given by
$$g_{\hh\hl\hl}={-3ig\mz\over 4\cos\theta\ls{W}}\sin 4\beta\,.\eqn\hhh$$
The rate obtained in eq.~\threeh\ is too small for a viable Higgs signal.

\FIG\hlbbbr{ The fractional deviation, $\Delta\br/
\br({\rm SM})$, is exhibited for $\hl\rta b\bar b$,
where $\Delta{\rm BR}\equiv {\rm BR(MSSM)}-{\rm BR(SM)}$ is the
difference between the corresponding branching ratios in the MSSM
and the Standard Model.  A given value of $\mha$ and
$\tanb$ determines the $\hl$ mass and couplings.
(Leading log radiative corrections have also been included, with
$m_t=175$~GeV, and a common mass of 1~TeV for all supersymmetric
particles.)}

\midinsert
   \tenpoint \baselineskip=12pt   \narrower
\centerline{\psfig{file=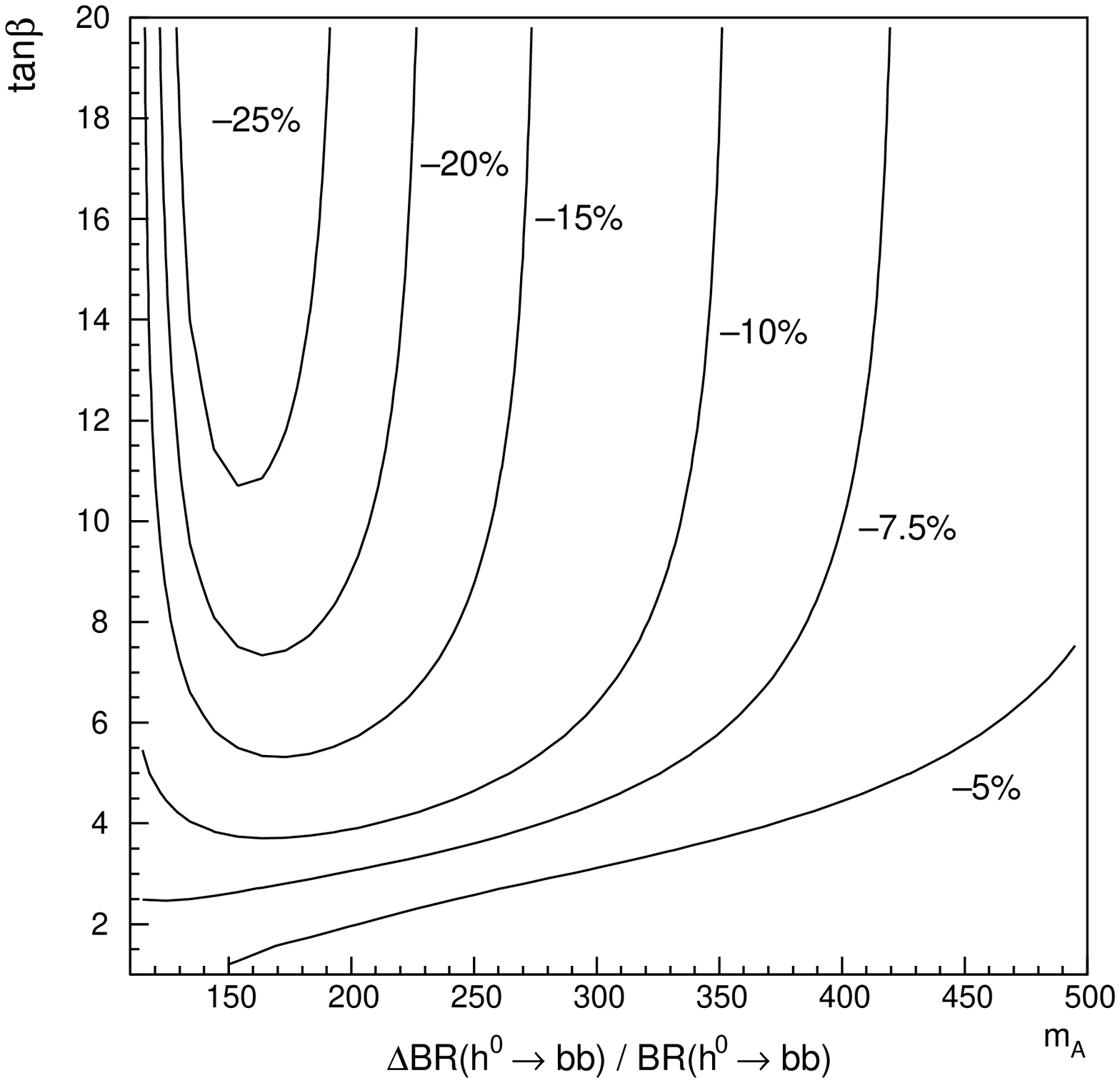,height=10cm}}
\vskip12pt\noindent
{\bf Fig.~\hlbbbr.}\enskip
Contour lines of fractional deviation, $\Delta\br/
\br({\rm SM})$, are exhibited for $\hl\rta b\bar b$,
where $\Delta{\rm BR}\equiv {\rm BR(MSSM)}-{\rm BR(SM)}$ is the
difference between the corresponding branching ratios in the MSSM
and the Standard Model.  A given value of
$\tanb$ and $\mha$ (the latter expressed in GeV units above)
determines the $\hl$ mass and couplings.
(Leading log radiative corrections have also been included, with
$m_t=175$~GeV, and a common mass of 1~TeV for all supersymmetric
particles.)
\endinsert

\REF\burke{M.D. Hildreth, T.L. Barklow, and D.L. Burke, {\sl Phys. Rev.}
{\bf D49} (1994) 3441.}
\REF\hildreth{H.E. Haber and M.D. Hildreth, in preparation.}
Second, consider precision measurements of $\hl$ branching ratios at
the NLC.  In a Monte Carlo analysis, Hildreth \etal\refmark\burke\
evaluated the
anticipated accuracy of $\hl$ branching ratio measurements at the NLC,
assuming $\sqrt{s}=500$~GeV and a data set of 50~fb$^{-1}$.   For
example, taking $\mhl=120$~GeV, they computed an extrapolated error
of $\pm 7\%$ for the $1$-$\sigma$ uncertainty in $\br(\hl\rta b\bar b)$
and $\pm 14\%$ for $\br(\hl\rta\tau^+\tau^-)$.  Other channels yielded
substantially larger uncertainties.  To see whether these are
significant measurements,
Hildreth and I compared the $\hl$ branching ratios
computed in the MSSM (including one-loop leading logarithmic radiative
corrections\refmark\impact) as a function of $\mha$ and $\tanb$
to the corresponding Standard Model branching
ratios.\refmark\hildreth\  (See ref.~\janot\ for a related analysis.)
The fractional deviation of $\br(\hl\rta b\bar b)$
from the corresponding Standard Model value is shown in Fig.~\hlbbbr.
[The analogous plot for $\br(\hl\rta\tau^+\tau^-)$ is nearly identical
to Fig.~\hlbbbr.]  Note that the (negative) deviation shown in Fig.~\hlbbbr\
is about 7\% for values
of $\mha$ as large as 450~GeV and about 15\% for values of $\mha$ as large
as 250~GeV.  These results
imply that a precision measurement of $\hl\rta b\bar b$ has the
potential for detecting the existence of a non-minimal Higgs sector
even if the heavier Higgs states cannot be directly detected at the
NLC.   Of course, one will have to push the precision of this
measurement beyond the present expectations to obtain a significant
result, since a 2-$\sigma$
deviation is not compelling evidence for new physics.  Other Higgs
decay channels are not competitive.

\FIG\hlgamgambr{ The fractional deviation, $[\br({\rm MSSM})-\br({\rm SM})]/
\br({\rm SM})$, is exhibited for $\hl\rta \gamma\gamma$.
See caption to Fig.~\hlbbbr.}

\midinsert
   \tenpoint \baselineskip=12pt   \narrower
\centerline{\psfig{file=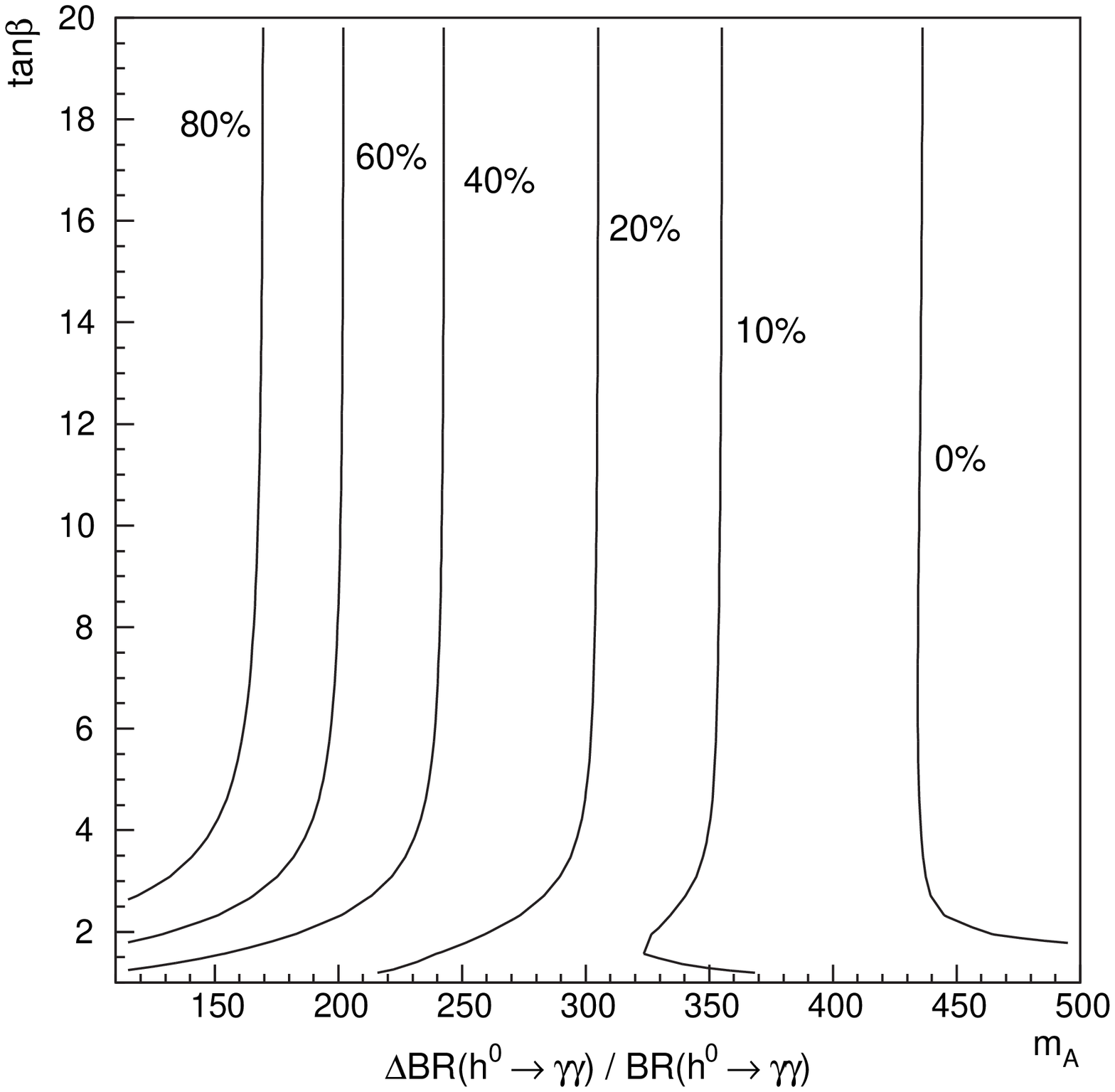,height=10cm}}
\vskip12pt\noindent
{\bf Fig.~\hlgamgambr.}\enskip
Contour lines of fractional deviation, $[\br({\rm MSSM})-\br({\rm SM})]/
\br({\rm SM})$, are exhibited for $\hl\rta \gamma\gamma$.
See caption to Fig.~\hlbbbr.
\endinsert
\medskip

\REF\gamgamcol{
H.F. Ginzburg, G.L. Kotkin, V.G. Serbo and V.I. Telnov,
{\sl Nucl. Inst. and Methods} {\bf 205} (1983) 47;
H.F. Ginzburg, G.L. Kotkin, S.L. Panfil,
V.G. Serbo and V.I. Telnov,
{\sl Nucl. Inst. and Methods} {\bf 219} (1984) 5.}

There is one novel approach which could extend the discovery limits
for heavy Higgs bosons at the NLC.  A high energy, high luminosity
photon beam can be produced by the Compton backscatter of an intense
laser beam off a beam of electrons.\refmark\gamgamcol\
This provides a mechanism for
turning the NLC into a high energy, high luminosity $\gamma\gamma$
collider.  All neutral Higgs states couple to $\gamma\gamma$ at one-loop
via loops of charged matter.
Since the couplings of the heavy Higgs
states to fermions and scalars are not suppressed in the decoupling
limit, the $\gamma\gamma$
couplings of the heavy Higgs states are also not suppressed relative to
the $\hl\gamma\gamma$ coupling.
Thus, one
can search for the non-minimal Higgs sector at the $\gamma\gamma$ collider by
either measuring the $\hl\gamma\gamma$ coupling with
sufficient precision or by directly producing $\ha$ and/or $\hh$ in
$\gamma\gamma$ fusion.  In the decoupling regime, the
$\hl\gamma\gamma$ coupling approaches the corresponding coupling of
the Standard Model Higgs boson, as illustrated in Fig.~\hlgamgambr.%
\foot{The contributions of
supersymmetric loops to the $\hl\gamma\gamma$ amplitude
vanish in the limit of large supersymmetric particle masses.}
As a result, this is not
a viable method for detecting deviations from the Standard
Model.  Thus, one must focus on $\gamma\gamma\rta(\ha,\hh)$.  In
ref.~\ghphoton, Gunion and I showed that parameter regimes exist where
one could extend the heavy Higgs mass reach above $\sqrt{s}/2$.
For example, at a 500~GeV $\gamma\gamma$ collider, a statistically significant
$\ha$ signal in $b\bar b$ and $Z\hl$ final states could be seen
above the background if $\mha<2\mt$.

\chapter{Conclusions}
\medskip

In the most general CP-conserving two Higgs doublet model, a
decoupling limit can be defined in which the lightest Higgs state is
a CP-even neutral Higgs scalar, whose properties approach those of the
Standard Model Higgs boson.
In the MSSM, the
decoupling limit corresponds to $\mha\gg\mz$ (independent of
$\tanb$).  Moreover, the approach to decoupling is rapid once $\mha$
is larger than $\mz$.  Thus, over a very large range of MSSM
parameter space, the couplings of $\hl$ to
gauge bosons, quarks and leptons are nearly identical
to the couplings of the Standard Model Higgs boson.

If the $\hl$ is discovered with properties approximating those of the
Standard Model Higgs boson, then the discovery of the non-minimal Higgs
sector will be difficult.  At the LHC, $\ha$, $\hh$ and $\hpm$
production rates via gluon-gluon fusion are not suppressed.  However,
isolating signals of the heavy Higgs states above background presents
a formidable challenge.  At the NLC, if $\sqrt{s}>2\mha$, then pair
production of $\hp\hm$ and $\hh\ha$ is easily detected.
Below pair-production threshold, detection of the non-minimal Higgs
sector is problematical.  For example, the
cross sections for single heavy Higgs boson production (in association
with light particles) are too small to be observed.  However,
experiments at the $\gamma\gamma$ collider may be able to
extend the NLC discovery limits of the heavy Higgs states (via
$\gamma\gamma$ fusion to $\hh$ or $\ha$).  Precision measurements of
$\hl\rta b\bar b$ may also provide additional evidence for a non-minimal Higgs
sector.  However, current experimental expectations at the NLC
predict only a 2-$\sigma$ deviation from Standard Model expectations
if $\mha$ is just beyond the kinematic limit
for pair production of heavy scalar states.

Once the first evidence for the Higgs boson is established, it will be
crucial to ascertain the underlying dynamics of the electroweak
symmetry breaking sector.  If the data reveals a single neutral CP-even
Higgs boson, with the precise properties expected of the minimal Higgs
boson of the Standard Model, then one will have to look for
alternative techniques of exploring the physics of
electroweak symmetry breaking.  The challenge for future collider
experiments is to develop new strategies for directly probing the
scalar sector in order to see beyond the minimal Higgs boson.


\medskip
\centerline{\bf Acknowledgments}
\medskip
This paper describes work performed in a number of separate
collaborations with Jack Gunion, Mike Hildreth,
Yossi Nir, and Scott Thomas.  Their contributions and insights are
gratefully acknowledged.
In addition, I would like to thank Jack Gunion
and his colleagues at the University of California, Davis
and Bernd Kniehl and his colleagues at the Max Planck Institute for Physics
in Munich for their warm hospitality and their labors in providing such
stimulating environments during their workshops.

\refout
\endpage
\bye